\documentclass[aps,pra,showpacs,amssymb,nofootinbib,superscriptaddress,twocolumn,longbibliography]{revtex4-1}
\usepackage{pstricks}
\usepackage{graphicx}
\usepackage{dcolumn}
\usepackage{bm}
\usepackage{dsfont}
\usepackage{amsmath}
\usepackage{pifont}
\usepackage{psfrag}

\usepackage{hyperref}

\usepackage{bbm}
\usepackage{amsthm}
\usepackage{ulem}
\usepackage{hyperref}
\hypersetup{colorlinks=true,urlcolor=blue,linkcolor=blue,citecolor=blue,filecolor=blue, linktocpage}

\usepackage{soul}

\newcommand{\BE}{\begin{equation}}
\newcommand{\EE}{\end{equation}}

\newcommand{\skipc}[2]{}

\newcommand{\fig}[1]{Fig.~\ref{#1}}
\newcommand{\eq}[1]{Eq.~(\ref{#1})}
\newcommand{\Sec}[1]{Sec.~\ref{#1}}
\newcommand{\App}[1]{App.~\ref{#1}}

\newcommand{\theo}[1]{Theorem~\ref{#1}}

\newtheorem{theorem}{Theorem}

\newcommand{\be}{\begin{equation}}
\newcommand{\ee}{\end{equation}}
\newcommand{\eea}{\end{eqnarray}}
\newcommand{\bea}{\begin{eqnarray}}

\newcommand{\ket}[1]{\ensuremath{|#1\rangle}}

\newcommand{\ketbra}[1]{\ensuremath{| #1 \rangle \!\langle #1 |}}

\newcommand{\Id}{\mathds{1}}

\usepackage{color}

\begin{document}

\title{Performance analysis of a hybrid agent for quantum-accessible reinforcement learning}

\author{Arne Hamann}
 \affiliation{Institut f\"ur Theoretische Physik, Universit\"at Innsbruck, Technikerstra{\ss}e 21a, 6020 Innsbruck, Austria}
 
\author{Sabine W\"olk}
 \affiliation{Institut f\"ur Theoretische Physik, Universit\"at Innsbruck, Technikerstra{\ss}e 21a, 6020 Innsbruck, Austria}
 \affiliation{Institute of Quantum Technologies, German Aerospace Center (DLR), D-89077 Ulm, Germany}

\date{\today}

\begin{abstract}
In the last decade quantum machine learning has provided fascinating and fundamental improvements to supervised, unsupervised and reinforcement learning. In reinforcement learning,  a so-called agent is challenged to solve  a task  given by some environment. The agent learns to solve the task by exploring the environment and exploiting the rewards it gets from the environment. For some classical task environments, such as deterministic strictly epochal environments, an analogue quantum environment can be constructed which allows to find rewards quadratically faster by applying quantum algorithms.
In this paper, we analytically analyze the behavior of a hybrid agent which combines this quadratic speedup in exploration with the policy update of a classical agent. This leads to a faster learning of the hybrid agent compared to the classical agent. We demonstrate that if the classical agent needs on average $\langle J \rangle$ rewards and $\langle T \rangle_c$ epochs to learn how to solve the task, the hybrid agent will take $\langle T \rangle_q \leq \alpha \sqrt{\langle T \rangle_c \langle J \rangle}$ epochs on average. Here, $\alpha$ denotes a constant which is  independent of the problem size. Additionally, we prove that if the environment allows for maximally $\alpha_o k_\text{max}$ sequential coherent interactions, e.g. due to noise effects, an improvement given by $\langle T \rangle_q \approx  \alpha_o\langle T \rangle_c/4 k_\text{max}$ is still possible.
\end{abstract}

\maketitle

\section{Introduction}
The application of quantum algorithms to machine learning provided promising results and evolved over the last years to the domain of quantum machine learning (QML) \cite{Dunjko2018,biamonte2017}. 

The main types of machine learning are supervised, unsupervised and reinforcement learning (RL) \cite{Sutton1998} and each of them can be improved by quantum algorithms \cite{biamonte2017,Dunjko2018, Ciliberto2017,Havlicek2019,Schuld2015,adcock2015advances}. In RL an agent has to solve a task via interactions with an environment, perceiving a reward as a measure of its performance on the task. RL can be applied to solve problems from different areas such as robotics \cite{johannink2019residual,tjandra2018sequence}, healthcare \cite{komorowski2018artificial}, or games such as Go \cite{Silver2016}.

The structure of RL allows for multiple quantum improvements. Various results show quantum advantages for quantum-enhanced agents interacting with a classical environment. In this way, improvements on the deliberation time of an agent \cite{Jerbi2021, PhysRevX.4.031002, Sriarunothai2018} or a better performance via variatonal quantum circuits \cite{jerbi2021variational} can be achieved.

Depending on the quantization of the environment, different methods can be applied \cite{Ronagh2019,Crawford2018,Cornelissen2018,Neukart2018,Casale2020,Saggio2021} to gain quantum improvements in sample complexity, that is the number of interactions the agent has to perform with the environment to solve the task. In this work, we will focus on an approach with quantum-accessible environments introduced  in \cite{Dunjko2016}. This framework allows for a quadratic speedup in sample complexity during the exploration on deterministic strictly epochal\footnote{Strictly epochal means, that consecutive interactions can be grouped into epochs of equal length and the environment is reset after each epoch.} environments. Generalizations of this framework  to stochastic environments \cite{Dunjko2016} and epochal environments with variable epochal length \cite{hamann2020quantumaccessible} also exist and for specific environments super-polynomial and exponential improvements have been demonstrated \cite{dunjko2017exponential}.

The quadratic speedup in exploration will improve the agent's performance if the agent is luck-favoring\footnote{\label{footnote:luckfavoring}Meaning that lucky agents, which received more rewards in the past, are expected to outperform unlucky agents, which received less rewards.} on the environment.  

In this paper, we investigate how this speedup in exploration will lead to an improvement in sample complexity. For this purpose, we introduce a hybrid agent based on a feedback loop between quantum exploration and classical updates. We then analyze the resulting performance and compare it to a similar classical agent based on the same update rules.

We determine possible speedups for two different situations: quantum-enhanced learning with (i) an  ideal quantum computer and (ii) a quantum computer with only a limited number of possible  coherent interactions.

An experimental implementation of such an agent with limited quantum resources based on a photonic quantum processor is described and demonstrated in \cite{Saggio2021}. In this paper, we concentrate on the theoretical background of such quantum-enhanced learning agents.

We will start with a brief review on reinforcement learning in \Sec{sec:rl} and  quantum-enhanced exploration in \Sec{sec:qrl}. Then, we will introduce a hybrid agent in \Sec{sec:agent} and analyze its behavior in \Sec{sec:analysis}. We conclude by summarizing our results and discussing generalizations of the here discussed hybrid agent to more general scenarios in \Sec{sec:conlusion}.


\section{Classical reinforcement learning\label{sec:rl}}

In reinforcement learning, an agent $A$ is challenged to solve a task via interactions with an environment $E$. The interaction is usually modelled as a (partially observable) Markov decision process (MDP) defined by the (finite) set of states $S$ of the environment and a (finite) set of actions $A$ performed by the agent.  

The  distribution of the initial state of the environment is described by a probability distribution $P(s)$.   An action $a$ performed by the agent leads to a state change of the environment from $s$ to $s'$  according to the transition probabilities $P(s'\vert a,s)$. The agent receives (partial) information, called observations or percepts $c(s)$, about the current state of the environment. The agent chooses its next action $a$ based on the observed percept $c(s)$ according to its current policy $\Pi(a\vert c(s))$. Additionally, the agent receives a real-valued reward $r$ rating its performance. The goal of the agent is to optimize the obtained reward in the long term by updating its policy $\Pi$ based on its observations and thus to learn. Different classical algorithms have been developed such as SARSA, Q-Learning, Deep Q-Learning or Projective Simulation \cite{Sutton1998,Mnih2015,Briegel2012} which provide good policies and update rules.

Some classical environments, such as deterministic strictly epochal (DSE) environments, can be transformed into a corresponding quantum environment, which allows to find rewards quadratically faster \cite{Dunjko2016}. Deterministic means that the initial state $s_0$ is fixed and the state $s'$ is completely determined by the preceding state $s$ and action $a$. In this case, we can assume that the  environment is fully observable such that the percepts $c(s)=s$ are equal to the states of the environment. 
 In strictly epochal environments, the interaction of an agent with its environment can be divided into epochs. Each epoch starts with the same initial state $s_0$ and consecutively $L$ percept-action pairs $(a_1,s_1),\cdots (a_L,s_L)$  are exchanged. 
As a consequence, the sequences of percepts $\vec{s}=(s_1,s_2,\cdots,s_n)$ and rewards $\vec{r}=(r_1,r_2,\cdots r_n)$ in a DSE environment are completely determined by the chosen sequence of actions $\vec{a}=(a_1,a_2,\cdots,a_n)$.

\subsection{The history of an agent}
The performed actions of an agent together with the resulting observed percepts and rewards form  its history
\begin{equation}
h_n=((s_1,a_1,r_1),\cdots,(s_{n-1},a_{n-1},r_{n-1}),(s_{n},a_{n},r_{n})).
\end{equation} 
An agent interacting with an environment will update its policy $\Pi(a\vert s)$ according to the observed history. Also the evaluation of the performance of an agent is usually a function based on its history.  

In general, the policy $\Pi(a\vert s)$ is probabilistic.  As a consequence, a set of learning agents defined by the same initial policy $\Pi_{h_0}$ and update rules  may observe different histories $h_n$ leading to different consecutive policies $\Pi_{h_n}$. We define by $H_n$ the set of all possible  histories $h_n$ with length $n$ which can be observed by this set of agents. The average performance of these agents thus depends on the probability distribution $p_n$ over the set of histories $H_n$. The probability $p_n(h_n)$ to observe a history of length $n$ can be determined  recursively. Let $h_{n+1}=((s_1,a_1,r_1),\cdots,(a_{n},s_{n},r_{n}),(a_{n+1},s_{n+1},r_{n+1}))$ be an extension of the history $h_n=((s_1,a_1,r_1),\cdots,(a_{n},s_{n},r_{n}))$. Then, the probability $p_{n+1}$ to observe the history $h_{n+1}$ is given by 
\begin{equation}
    p_{n+1}(h_{n+1}) = \Pi_{h_n}\left(a_{n+1}|s_{n+1}\right)\cdot P\left(s_{n+1}|a_n,s_n\right)\cdot p_n(h_n),
\end{equation}
where $\Pi_{h_n}$ denotes the actual policy of an agent with observed history $h_n$. The recursion starts with $p_0(h_0)=1$ and 
$P(s_1|a_0,s_0) = P(s_1)$.

In order to solve a given task, agents with different histories  usually need a different number of interactions $n$. Therefore, it is necessary to extend the probability distribution $p_n(h_n)$ over the set $H_n$ of histories with length $n$ to the probability distribution $p(h_n)$ over the set of infinite histories  $H^\infty$ \cite{maschler2020game}. Here, $p(h_n)$ now determines the probability  that an infinite long history $h_\infty$ starts with $h_n$ via
\begin{equation}\label{equ:p}
     p(h_n):= p(\{h_\infty\vert h_\infty \text{ starts with }h_n\}) =p_n(h_n).
\end{equation}

For simplicity and readability we will for now on use the shorthand notation $p(h)$ introduced in \eq{equ:p}. 

In \Sec{sec:analysis}, we will use the probability distribution of histories $p(h_n)$  to compare a set of classical and quantum-enhanced agents in order to determine possible quantum speedups. To derive analytical results, we will concentrate on classical learning agents which fulfill the following two criteria:  First, policy updates are completely determined by the reward history, that is the history of the agent reduced to  epochs, where a non-vanishing reward was obtained. Second, the agents plans its actions for the complete next epoch at the beginning of this epoch. This can be achieved e.g. by using mapping as outlined in the next section. In this way, it is possible to define  probabilities $\hat{\Pi}(\vec{a})$ for all possible action sequences $\vec{a}$ for the next epoch.

\subsection{Mapping\label{sec:model}}
\begin{figure}
    \centering
    \includegraphics[width=0.4\textwidth]{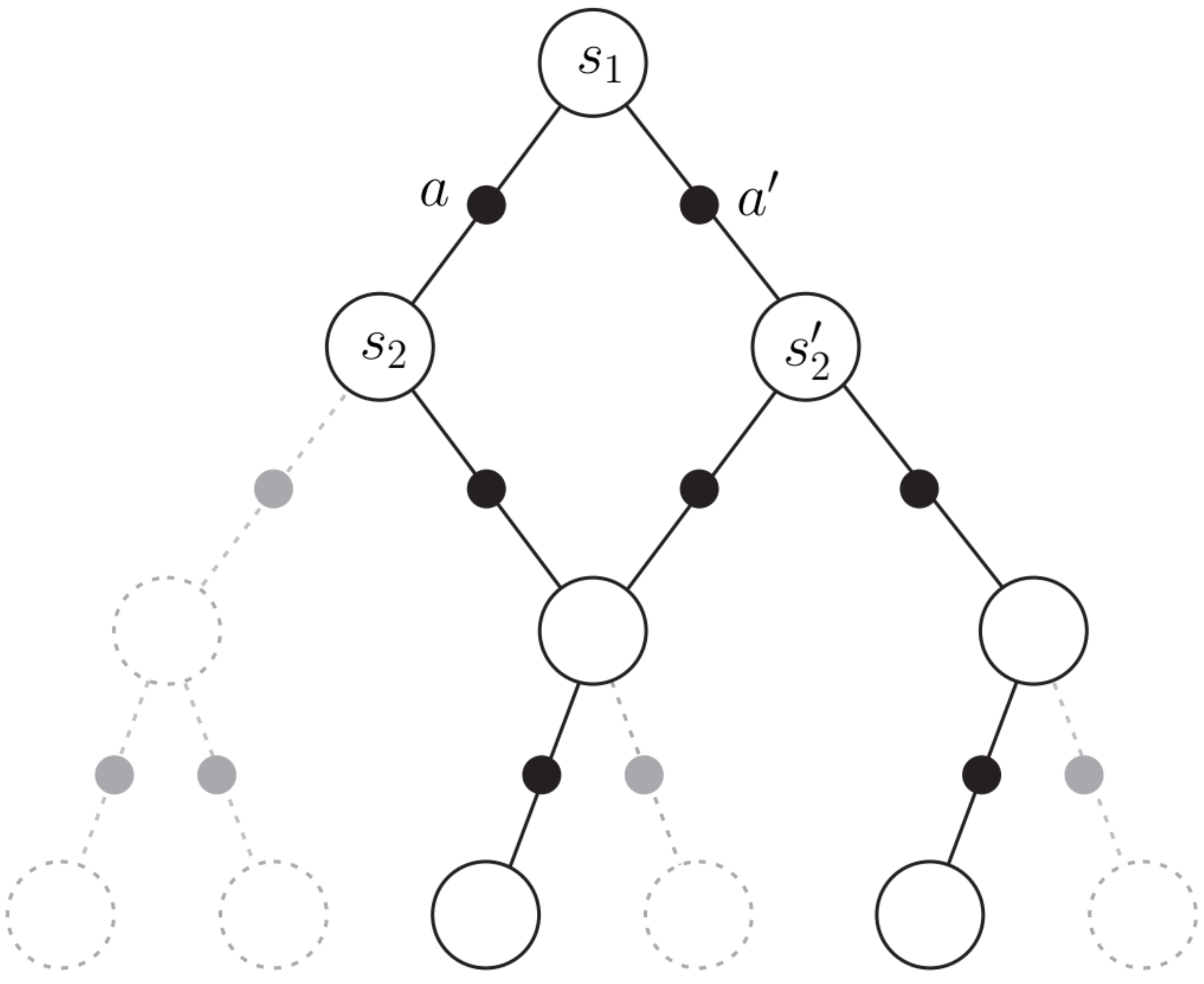}
    \caption{Map generated by an agent interacting with a deterministic strictly epochal environment containing all observed percepts $s$ (open circle), actions $a$ (small filled circle) and transitions (lines). It assumes that  actions not performed so far lead to new percepts (grey, dashed).   }
    \label{fig:decision_tree}
\end{figure}

In general, learning agents interacting with a DSE environment can also use mapping or model building \cite{Sutton1998} in order to determine their next actions. A map of the environment can e.g. be generated by not only generating and storing a policy $\Pi(a\vert s)$ for every observed percept $s$, but also keeping track of  observed transitions $(s_n,a_n) \longrightarrow s_{n+1}$. Such a map can also include actions and percepts not taken/observed so far. \fig{fig:decision_tree} shows an example of a map where each epoch consists of 3 consecutive actions and in each step, the agent can choose between two different actions $a$ and $a'$. Here, the agent already observed that starting an epoch with the sequence of actions $(a_1=a,a_2=a')$ leads to the same percept as starting with $(a_1=a',a_2=a)$. However, so far it has  not observed e.g. the percept resulting from starting the epoch with the action sequence $(a_1=a,a_2=a)$. Therefore, the agent assumes that this sequence of actions will lead to a new, so far unobserved percept.

An agent can use such a map to plan its action for the next epoch by defining the policy
\begin{eqnarray}
\hat{\Pi}(\vec{a})&=&\prod\limits_{j=1}^L \Pi(a_j\vert s(a_1,\cdots,a_{j-1}))
\end{eqnarray}
for complete action sequences of an epoch. Here, it uses
for unknown percepts $s$ the equal distribution
\begin{eqnarray}
\Pi(a_j\vert s(a_1,\cdots,a_{j-1}))&=&\frac{1}{||A||}\quad \text{if }  s \text{ unknown}, 
\end{eqnarray}
where $||A||$ determines the number of possible actions.  

The above described method is only one possible way how an agent can plan its actions for the next epoch. The important point is that it is possible to define $\hat{\Pi}(\vec{a})$ at the beginning of each epoch. 
In \Sec{sec:agent}, we will combine such a classical agent, which plans its actions at the beginning of each epoch,  with quantum amplitude amplification in order to create a quantum-enhanced learning agent.


\section{Quantum-enhanced exploration\label{sec:qrl}}
To quantize RL we will use the approach of quantum accessible environments introduced in \cite{Dunjko2016}, where the agent and the environment are embedded in a communication scenario. The environment sends percepts  and rewards  to the agent, which responds with actions. This communication is quantized by encoding classical percepts, actions and rewards into orthonormal quantum states. In this scenario, the quantum-enhanced agent can choose to interact quantum by using superposition states as action states or classical by limiting itself to orthonormal basis states.

In DSE environments as considered in this paper, the sequence of observed percepts $\vec{s}$ and rewards in an epoch are completely determined by the chosen sequence of actions $\vec{a}$ within this epoch. In addition, the length of all epochs is fixed and at most one reward $r$ is assigned. Thus, there exist a unitary
\BE
U_\text{Env}\ket{\vec{a}}_A\ket{0}_S\ket{0}_R = \ket{\vec{a}}_A\ket{\vec{s}(\vec{a})}_S\ket{r(\vec{a})}_R. \EE
describing the overall effect of the environment within one epoch.

In general, this leads to entanglement between the action, percept and reward register if the action register is in a superposition state. This entanglement can cause decoherence of the action state when the percept and reward states are reinitialized for the next epoch. However, the entanglement between actions and percept states can be coherently disentangled for
deterministic strictly epochal environments with the help of  a second epoch (for details see Ref. \cite{Dunjko2016}), such that the overall effect of the environment is equal to an oracle described by the unitary 
\BE
O_E \ket{\vec{a}}_A \ket{0}_S\ket{0}_R = \left \lbrace \begin{array}{cr}\ket{\vec{a}}_A \ket{0}_S\ket{0}_R & \text{if } r(\vec{a})=0\\
\ket{\vec{a}}_A \ket{0}_S\ket{1}_R & \text{if } r(\vec{a})>0 \end{array}\right. .
\EE

The oracle $O_E$ is equivalent to a controlled not-gate acting on the reward register and controlled by the action state $\ket{\vec{a}}_A$. It is possible to build a quantum-enhanced learning agent as described here, if the effect of the environment can be transformed into such an oracle by playing a single ($\alpha_o=1$, e.g. for action independent percepts) or several epochs ($\alpha_o=2$, e.g. for deterministic strictly epochal environments).

Quantum-enhanced agents using this oracle can find rewarded sequences of actions quadratically faster. That is, for a fixed policy, they need on average $\langle t \rangle_q = \alpha \sqrt{\langle t \rangle_c} $ \cite{Grover1998, Boyer1998, PhysRevLett.113.210501} epochs to find the next reward, whereas a classical agent would need $\langle t\rangle_c$ epochs. Here, the constant $\alpha=\alpha_s\alpha_o$ is determined by the number of epochs $\alpha_o$ necessary to create the oracle $O_E$ and a constant $\alpha_s$ depending on the applied quantum search algorithm \cite{Grover1998, Boyer1998, PhysRevLett.113.210501}\footnote{Typical values are $\alpha_s=\frac{\pi}{4}$ if the classical reward probability is known  or $\alpha_s=\frac{9}{4}$ if it is unknown.}. 
Given this quadratic speedup in exploration, it is possible to construct  a basic quantum agent based on any classical agent. A basic quantum agent performs quantum searches for a certain amount of time and then trains an internal copy of the classical agent to reproduce the found rewarded sequences of actions. This basic quantum agent is on average luckier than the classical agent, as it will on average find rewarded sequences faster than the classical agent. Hence, if the agent-environment-setting is luck-favoring[\ref{footnote:luckfavoring}], it will outperform the classical agent\cite{Dunjko2016}.

However, more advanced quantum-classical hybrid agents can be constructed by alternating between quantum search and classical policy updating as discussed in this paper and experimentally demonstrated in \cite{Saggio2021}.  Furthermore, we quantify the overall speedup in learning for these agents, which is in general not possible for the basic quantum agent described in \cite{Dunjko2016}.


\section{Description of the hybrid learning agent\label{sec:agent}}

\begin{figure*}
\begin{center}
\includegraphics[width=0.8\textwidth]{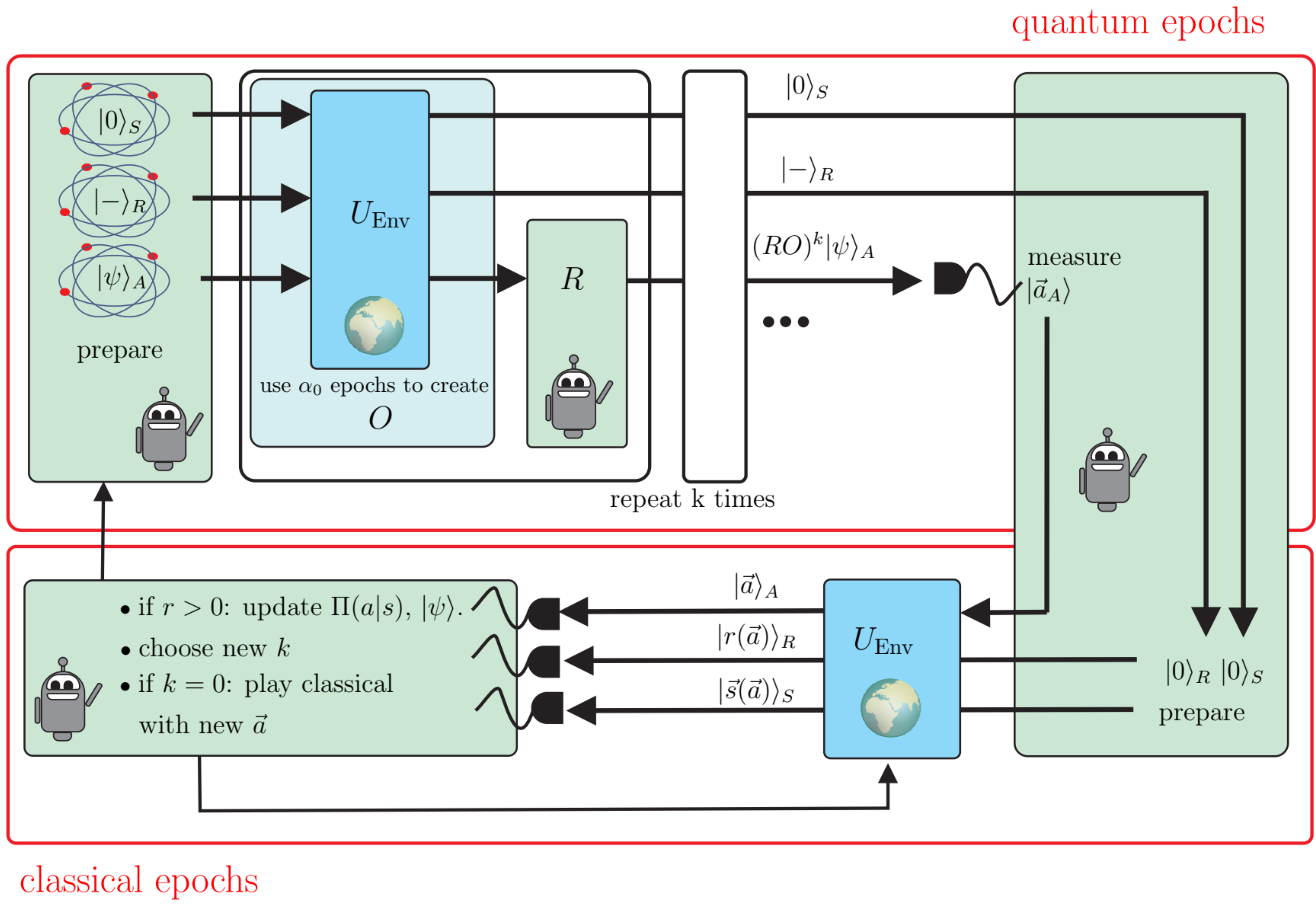}
\caption{\textbf{Scheme of a hybrid learning agent:} A hybrid agent alternates between quantum and classical epochs. A sequence of  quantum epochs is started by preparing the reward register  and the action register in the superposition states $\ket{-}_R$ and $\ket{\psi}_A$. By playing $\alpha_o k$ epochs, the actions state is transformed into $(RO)^k\ket{\psi}$ whereas the final reward and percept register are equal to their initial states. Consecutively, the agent performs a measurement on the action register resulting in a single sequence of actions $\vec{a}$. Subsequently, the agent plays one classical epoch where its actions are described by the measurement result $\vec{a}$
. The agent uses the obtained information about percepts and rewards from the classical epoch to update its policy. Based on the updated policy, the agent decides to either continue with quantum or classical epochs.  }
\label{fig:agent}
\end{center}
\end{figure*}

A key feature of reinforcement learning is the assumption that there exist not only correct and incorrect (sequences of) actions  but rather a spectrum of better or worse (sequences of) actions. In addition, some resemblance  between good (sequences of) action is usually implied. In these cases, finding good actions sequences which might be suboptimal (but with higher probability to find them) can help to find better (sequences of) actions, which are less probable. For example, there may exist many, not too long routes from city $A$ to city $B$ with different length. By slightly varying such a route, an even shorter and therefore better route might be found. As a consequence, reinforcement learning problems with DSE environments can in theory be solved by testing all possible action sequences and searching for the optimal one. However, the search space for such problems is usually too big and the optimum might be unknown such that solving the problem via straight forward search is not possible in practice and reinforcement learning is used instead. 

Quantum-enhanced exploration as discussed in \Sec{sec:qrl} can speed-up the search for rewards quadratically. However, the search space might  nevertheless be too large to find the optimal solution via direct search. We therefore use in these cases quantum-enhanced exploration to search for general rewarded actions and use classical methods for policy updates. The policy updates will change  the underlying search space and help to find better solutions. In general, also these modified search spaces are  still big such that quantum-enhanced exploration can be also advantageous after the first policy updates. Therefore, we introduce in the following a hybrid quantum-classical learning agent. This agent  uses quantum-enhanced exploration not only in the beginning, before any policy update took place. Thought, it alternates between quantum-enhanced exploration and classical policy updates. We establish a feed-back loop between quantum-enhanced exploration and classical policy updates such that both procedures can profit from each other. 

After describing our quantum-enhanced learning agent, we will compare in the next section its behavior with a corresponding classical agent based on the same initial policy and policy update rules.  

In the following, we consider classical agents with a policy which can be described by a
 a probability distribution $\hat{\Pi}(\vec{a})$ for action sequences for a complete epoch. This is e.g. possible for agents in DSE environments which use mapping as described in \Sec{sec:model}. 

We can create a hybrid quantum classical learning agent based on this classical agent. The hybrid agent alternates between quantum epochs used for quantum exploration and classical epochs used for policy updates as shown in \fig{fig:agent} and described in detail below:

\begin{enumerate}
\item Given the classical probability distribution $\hat{\Pi}(\vec{a})$ of action sequence, estimate a lower bound $Q_{min}$ on the winning probability
\begin{equation}
Q=\sum\limits_{\lbrace \vec{a}|r(\vec{a})>0\rbrace}\hat{\Pi}(\vec{a})\label{eq:win_prob}
\end{equation}
and prepare the action state 
\begin{equation}
\ket{\psi}_A=\sum\limits_{\lbrace \vec{a}\rbrace}\sqrt{\hat{\Pi}(\vec{a})}\ket{\vec{a}}_A
\end{equation}
and the reward state $\ket{-}_R=(\ket{0}_R-\ket{1}_R)/\sqrt{2}$.
\item Use several epochs to perform  amplitude amplification \cite{Boyer1998, Grover1998} until a reward is found. This consists e.g. of the following steps, if the winning probability $Q$ is not known exactly:
\begin{enumerate}
\item Initialize $m=1$ and $\lambda=6/5$ and choose a random integer $k<m$.
\item Use $\alpha_o k$ epochs to perform $k$ Grover iterations 
\begin{equation}
\ket{\psi'}_A\ket{-}_R=\left(R O_E\right)^k\ket{\psi}_A\ket{-}_R,
\end{equation}
where $R = \Id - 2 \ketbra{\psi}$ is the reflection around the initial action state.
\item Measure the resulting $\ket{\psi'}_A$ to determine a possible sequence of actions $\vec{a}$.
\item Play one classical epoch with the measured sequence of actions $\vec{a}$ and record the observed percepts and reward.
\item Terminate if the sequence wins a reward; else set $m$ to min$(\lambda m, \sqrt{1/Q_{min}})$ and restart.  
\end{enumerate}
\item Use the most recent classical information from step 2(d) to update the classical policy.
\item Determine the new probability distribution $\hat{\Pi}(\vec{a})$ and $Q_{min}$ according to the new policy and repeat.
\end{enumerate} 

The probability to observe a reward in step 2(d) after performing $k$ Grover iterations and playing $\alpha_o k+1$ epochs is given by (see \App{appendix:Q_max})
\be
G(Q,k)=\sin^2\left[(2k+1)\arcsin(\sqrt{Q})\right].\label{eq:G}
\ee
A classical agent sampling its actions  directly from $\hat{\Pi}(\vec{a})$ would observe a reward with probability $Q$ in each single epoch. As a consequence, performing Grover iterations leads to observing a reward more frequently only if $Q<G(Q,k)/(\alpha_o k+1)$.
This is only the case for winning probabilities $Q$ below a certain threshold $Q\leq Q_\text{max}$. This threshold is given by $ Q_\text{max}\approx 0.3964$ for simple learning problems with $\alpha_o=1$ and the minimal number of Grover iterations $k=1$. 

A hybrid agent following the steps 2(a)-(e) will automatically decrease the probability to perform Grover iterations because it always starts with $k=0$ each time after it has found a reward. Thus performing Grover iterations becomes more and more unlikely the larger $Q$. Nevertheless, it might be advantageous to fix $k=0$, and thus restrict the agent to classical behavior, at a certain point of learning. This behavior can be steered e.g. by the total number of found rewards or the observed frequency of rewards in the last few classically played epochs.


\section{\label{sec:analysis}Analysis of the hybrid agent}

\begin{figure*}
    \centering
    \includegraphics[width=0.7\linewidth]{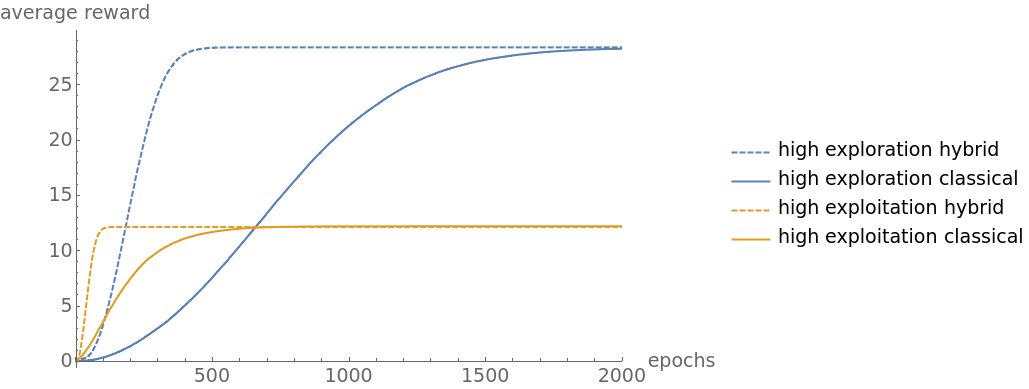}
    \caption{Simulation of the average reward of hybrid and  classical learning agents based on t-PS \cite{Flamini_2020} on a binary tree environment with 12 layers ($2^{12}$ leaves and $2^5$ rewarded leaves). 
    The maximum reward, given by $r=2^5$, is obtained on the optimal path through the tree. 
    Leaves reached by leaving the optimal path after $x$ decisions are rewarded with $r=\lfloor2^{(5+x-12)}\rfloor$. Shown is the average reward of 50000 agents over time for more explorative agents with $\beta=0.01$ (blue curve) and more exploitative agents with $\beta=0.1$ (orange curve).
    Further details such as the definition of $\beta$ can be found in \App{appendix:binary_tree_env}.
    }
    \label{fig:learning-curve}
\end{figure*}

In the following, we compare the behavior of the above described hybrid learning agents with corresponding classical agents. The behavior of a learning agent depends on the history it has observed. Hybrid agents only observe percepts and rewards during classical epochs (step 2(d)). As a consequence, the general history of a hybrid agent, containing all actions, percepts and rewards for all epochs, is not well defined. However, we can define the rewarded history $h_r$, which is the history of an agent reduced to rewarded epochs. That is, epochs where a non-vanishing reward $r>0$ was found. E.g. when a non-vanishing reward was obtained in every even epoch, the corresponding rewarded history would be $(e_2,e_4,e_6,...)$ with $e_j$ containing the sequence of actions, percepts and rewards observed in epoch $j$. We define  the event of observing a rewarded history $h_r$ as the set of all infinite length histories, which reduce to rewarded histories starting with $h_r$. 

Therefore, we consider only agents which behavior is completely defined by their rewarded history. That is, their behavior can be described by    
 $\hat{\Pi}(\vec{a})$ and they only update their policy (and thus also their map of the environment) when receiving a reward. In addition, all agents (hybrid and classical) use the same update rules which are based solely on rewarded epochs. As a consequence, the behavior of all agents  is solely determined by rewarded epochs. 

The analysis of the hybrid agent will concentrate on three properties. First, we will show that the probabilities to observe a given rewarded history are identical for the classical and the corresponding hybrid agent. Then, we will focus on the scaling advantage in learning time of the hybrid agent with unlimited quantum resources. In the end, we will go one step into the direction of  Noisy Intermediate-Scale Quantum (NISQ) computers and investigate achievable improvements based on a limited number of coherent interactions.

\subsection{Distribution of rewarded histories $h_r$}

The final quality of a trained agent depends on the initial policy and performed policy updates and thus in the here considered cases solely on the observed rewarded history. That is the history reduced to rewarded epochs. As a consequence, a classical agent and a hybrid agent which have observed the same rewarded history $h_r$ obtain the same classical policy. In addition, the probability $p(h_r)$ to observe a given rewarded history $h_r$ is equal for a hybrid agent and its corresponding classical agent as stated below and proven in \App{appendix:same_rewarded_histories}:
 
\begin{theorem}\label{theorem:same_rewarded_histories}
Let $E$ be a deterministic (time independent) strictly epochal environment and $O_E$ be the corresponding reward oracle and $A_c$ be a corresponding  classical agent to the hybrid agent $A_q$ as defined above. Then, the probability distribution of rewarded histories $p(h_r)$ for $A_c$ interacting with $E$ and $A_q$ interacting with $O_E$ and $E$ are equal.
\end{theorem}

\theo{theorem:same_rewarded_histories} leads to several direct consequences when comparing a group of trained hybrid agents with a group of corresponding classical agents. 

Consider for example a setup with different possible rewards $r$. Not all agents will learn to achieve the maximal possible reward in the long run due to their individual observed histories. Yet,  
 a group of hybrid agents and a group of corresponding classical agents converge towards the same average reward as shown in \fig{fig:learning-curve} due to \theo{theorem:same_rewarded_histories}. 
 
 Or consider groups of agents, where each agent plays until it has found $J$ rewards. Then, all hybrid agents switch to complete classical behavior and we compare the consecutive behavior of the group of hybrid agents  with the group of classical agents. In this case, both groups behave exactly similar and the behavior of both groups  are indistinguishable from each other.

In ML, there often exists a trade-off between exploration and exploitation.  This manifests itself often in a trade-off between fast learning and optimal behavior 
in the long run\cite{Sutton1998,MelnikovMakmalBriegel2018}.  
An example of such a trade-off is visualized in \fig{fig:learning-curve}. Here, we considered classical policies depending on some parameter $\beta$ influencing the ratio between exploration and exploitation (see \App{appendix:binary_tree_env}). Exploitative agents ($\beta=0.1$, orange curve) learn faster but get stuck more likely in local maxima. As a consequence, the expected average reward will be lower in the long run. Whereas explorative agents ($\beta=0.01$, blue curve) learn slower but reach in the long run a higher expected average reward. This is the case for classical agents as well as for hybrid agents.  

In contrast, going from a classical agent to its corresponding hybrid agent, both based on the same policy, leads to faster learning without sacrificing the expected average reward due to \theo{theorem:same_rewarded_histories}.


\subsection{Learning time}
In order to quantify the speedup of our hybrid agent  we define the learning time $T(h)$ of an agent with history $h$ as the minimal number of epochs $t$ this agent needs to reach a winning probability $Q=Q_t$, \eq{eq:win_prob} above a predefined learning threshold that is $Q_t  \geq Q_l$.
We say the learning time is infinite if $Q_t < Q_l$  $\forall t$.
A learning threshold is achievable, if and only if the probability $p(\{h_\infty\in H_\infty|T(h_\infty)=\infty \})=0$ for histories of agents with infinite learning time is zero. 

In general, a speedup in learning is only achieved while the hybrid agent performs amplitude amplification which is only the case for winning probabilities $Q<Q_\text{max}$ (compare \Sec{sec:agent}). In these cases,  our hybrid agent can achieve the following speedup compared to its corresponding classical agent: 
\begin{theorem}\label{theorem:learning_time}
Let $E$ be a deterministic (time independent) strictly epochal environment and $O_E$ be the corresponding reward oracle.
Then, for all achievable reward probabilities $Q_l<Q_\text{max}$, 
the average learning time of a hybrid agent $\langle T_q \rangle$ and its corresponding classical agent $\langle T_c\rangle$ are connected via\footnote{Typical $\alpha_{s,o}$ are of the order $\alpha_o \in \{1,2\}$ and $\alpha_s \approx \frac{9}{4}$}
\be
\langle T\rangle_q \leq \alpha \sqrt{\langle T \rangle_c \langle J \rangle}
\ee
where $\langle J\rangle$ denotes the average number of rewards  agents need to observe in order to learn.
\end{theorem}
Hence the hybrid agent learns faster, while it converges towards the same average policy as its corresponding classical agent.

\begin{proof}
In order to determine the average $\langle T\rangle$, we split the learning time into intervals of length $t_j$ of constant policy. Thus, the interval $j$ starts after $j$ non-vanishing rewards have been observed and ends with the observation of the next non-vanishing reward. In addition, we define $J(h)$ as the number of non-vanishing rewards an agent with history $h$ has observed until it has learned. As a consequence, the learning time of an agent with history $h$ is given by
\be
\label{equ:T} T(h)=\sum_{j=0}^{J(h)-1} t_j(h).
\ee 
The average learning time $\langle T \rangle$ is thus determined by averaging $T(h)$ over all possible histories $h$. We perform this averaging in two steps. First, we average over all histories $h$ which can be reduced to the same rewarded history $h_r$. Consecutively, we average over all rewarded histories $h_r$.

The policy $\hat{\Pi}$, the winning probability $Q$ and thus $J$ depend not on the exact history $h$ but solely on the rewarded history $h_r$ for the learning agents considered here. Thus, an agent which has found already $j$ rewards with rewarded history $h_r$ needs on average
\begin{equation}
    \langle t_j(h_r) \rangle_c = \frac{1}{Q_j(h_r)} \label{eq:t_j(h)}
\end{equation} 
epochs to find the next reward (see \App{appendix:proof_learning_time} for a more detailed discussion).
A quantum-enhanced agent as described in \Sec{sec:agent} finds rewards quadratically faster \cite{Boyer1998, Grover1998}. As a consequence, the average interval time  $\langle t_j(h_r) \rangle_q $ for such agents is given by 
 \begin{equation}
 \langle t_j(h_r) \rangle_q \leq \frac{\alpha}{\sqrt{Q_j(h_r)}}= \alpha \sqrt{\langle t_j(h_r) \rangle_c}.
\end{equation}

The learning time $\langle T(h_r)\rangle $ averaged over all agents with the same rewarded history $h_r$ is determined by 
\begin{equation}
    \langle T(h_r) \rangle  = \sum_{j=0}^{J(h_r)-1} \langle{t_j}(h_r)\rangle.
\end{equation}
As a consequence, quantum-enhanced learning agents with a rewarded history $h_r$ learn on average in
\begin{eqnarray}
    \langle T(h_r) \rangle_q  &\leq& \sum_{j=0}^{J(h_r)-1} \alpha \sqrt{\langle t_j(h_r) \rangle_c}\\
    &\leq& \alpha \sqrt{J(h_r)} \sqrt{\langle T(h_r) \rangle_c}
\end{eqnarray}
epochs. Here, we have used the Cauchy-Schwarz inequality in the second step. 

Averaging $\langle T(h_r)\rangle$ over all possible rewarded histories leads to (see \App{appendix:proof_learning_time})
\be
\langle T\rangle =\sum\limits_{\lbrace h_r\rbrace} p(h_r)\langle T(h_r)\rangle.\label{eq:T_average}
\ee
Note, that the probability $p(h_r)$ to observe a given rewarded history is identical for quantum-enhanced learning agents and their corresponding classical agents  due to \theo{theorem:same_rewarded_histories}. As a consequence, using again the Cauchy-Schwarz inequality leads to
\be
\begin{split}
\langle T\rangle_q&\leq \alpha\sum\limits_{h_r} p(h_r)\sqrt{J(h_r)\langle T(h_r)\rangle_c} \\
&\leq \alpha\sqrt{\langle J \rangle \langle T \rangle_c},
\end{split}
\ee
where $\langle T\rangle_q$ and  $\langle T\rangle_c$ denote the average learning times of the quantum-enhanced agents and classical agents, respectively. $\langle J\rangle$ denotes the average number of rewards an agent with given policy update rules needs to find in order to learn.
\end{proof}


\subsection{Noisy quantum devices}
\theo{theorem:learning_time} quantifies the achievable speedup of our hybrid agent assuming the existence of a perfect quantum computer. However, quantum computers which will be available soon  will be noisy and thus possess limited coherence times. Such behavior can be approximated by assuming that noise can be neglected for up to $\alpha_o k_\text{max}$ consecutive quantum epochs whereas the effect of noise starts to become crucial if more than $\alpha_o k_\text{max}$ quantum epochs are performed consecutively. Therefore, the question arises which improvements can be achieved if only a limited number of epochs can be performed coherently in a row. That is, we investigate the achievable improvement assuming that the  number of Grover iterations $k$ in step 2(b) of our hybrid agent is limited to $k\leq k_\text{max}$. 
Obviously, the limitation of $k$ plays only a role if the winning probability $Q$, \eq{eq:win_prob}, is small such that
\be
Q\leq Q_{k_\text{max}}=\sin^2\left[\frac{\pi}{2(2k_\text{max}+1)}\right].
\ee
Here, $Q_{k_\text{max}}$ denotes the smallest reward probability which leads to $G(Q_{k_\text{max}},k_\text{max})=1$, see \eq{eq:G}, when performing $k_\text{max}$ Grover iterations.
It is also possible to achieve an improvement in this regime albeit a linear one as stated in the following theorem:

\begin{theorem}\label{theorem:learning_time_restricted}
Let $E$ be a deterministic (time independent) strictly epochal environment and $O_E$ be the corresponding reward oracle. 
Then, a hybrid learning agent as defined above and limited to maximal  $k_\text{max}$ sequential Grover iterations interacting with $O_E$  can reach  achievable reward probabilities $Q_l<Q_{k_\text{max}}$ in a learning time  $\langle T \rangle_q$ with 
\be
\langle T\rangle_q \leq \frac{\alpha_o\pi^2}{16}\frac{\langle T \rangle_c}{k_\text{max}}, 
\ee
where $\langle T\rangle_c$ is the learning time of the corresponding classical agent interacting with $E$.
\end{theorem}
Note, this is a linear improvement in query complexity which can be crucial in certain settings independent of the running time of the algorithm.
\begin{proof}
Again, we first consider a learning agent with a given rewarded history $h_r$ before we average over all possible rewarded histories.

A classical winning probability $Q_j=Q_j(h_r)$ leads after $k_\text{max}$ Grover iterations to an enhanced winning probability  \cite{Boyer1998, Grover1998}
\be
G(Q_j,k_\text{max})=\sin^2\left[(2k_\text{max}+1)\arcsin (\sqrt{Q_j})\right].
\ee
As a consequence, the expected interval time $t_j$ (compare proof 
of  \theo{theorem:learning_time})  of the hybrid agent is given by
\be
\langle t_j(h_r)\rangle_q=\frac{\alpha_o k_\text{max}+ 1}{G(Q_j,k_\text{max})}.\label{eq:T_noise}
\ee
Here, we have taken into account that $\alpha_o k_\text{max} $ epochs are necessary to create $k_\text{max}$ Grover iterations plus one epoch to determine the reward.
The denominator can be approximated with the help of the inequality
 $\sin(x)\geq \beta x$, which is valid in a range $0\leq x \leq x_0(\beta)$ determined by $\beta$. In our case, we consider the interval $0\leq x\leq \frac{\pi}{2}$ and therefore need to chose $\beta=\frac{2}{\pi}$.

Identifying $x=(2k_\text{max}+1)\arcsin(\sqrt{Q_j})$ and using $\arcsin(\sqrt{Q_j}) >\sqrt{Q_j}$ leads to 
\be
\begin{split}
G(Q_j,k_\text{max}) &\geq \frac{4}{\pi^2}(2k_\text{max}+1)^2  Q_j\\
&\geq \frac{4}{\pi^2} 4k_\text{max}(k_\text{max}+1) Q_j. 
\end{split}
\ee 
With the help of $ \langle t_j(h_r)\rangle_c=1/Q_j$ and $1\leq \alpha_o$, we find as a result
\be
\langle t_j(h_r)\rangle_q \leq \frac{\alpha_o\pi^2}{16k_\text{max}} \langle t_j\rangle_c.
\ee
A summation over $0  \leq j< J(h_r)$ gives the average learning time $\langle T(h_r) \rangle$ and the average over all possible rewarded histories $h_r$ leads to \eq{eq:T_noise} due to \theo{theorem:same_rewarded_histories}.
 \end{proof}
 Notice that the inequality can be tightened by using bigger values for $\beta$, allowing to prove higher improvements for smaller limits on $Q_l$. The extreme case with $\beta=1$ is approximately reachable for $Q_l\ll1$ and $k_\text{max}\ll \frac{1}{Q_l}$ leading to \be \langle T_h\rangle \approx \alpha_o \frac{\langle T_c \rangle}{4 k_\text{max}} .\ee
 
 In general, the total learning process of a quantum-enhanced learning agent with limited quantum resources can be split into three phases. The first phase is defined by winning probabilities $Q_j \leq Q_{k_\text{max}}$. In this phase,  a linear improvement proportional to  $k_\text{max}$ is achievable according to \theo{theorem:learning_time_restricted}. The second phase is defined by $Q_{k_\text{max}}< Q_j < Q_\text{max}$. Here, a complete Grover search is possible and beneficial and a quadratic speedup is achieved according to \theo{theorem:learning_time}. The last phase $Q_j\geq Q_\text{max}$ is the phase, where the hybrid agent reproduces the classical agent and therefore no speedup can be generated in this phase.

\section{Conclusions and Outlook\label{sec:conlusion}}

In this paper, we analyzed the quantum speedup which can be gained by combining classical reinforcement learning agents with quantum exploration. We compared quantum-enhanced learning agents, which alternate between quantum exploration and classical policy updates, with classical learning agents based on the same policies and update rules.

For analytical reasons, we considered only agents which fulfilled the following two criteria: First, agents are able to determine the probabilities for all possible action sequences for one epoch beforehand. Second, policy updates are completely determined by the rewarded history, that is the history of an agent reduced to  epochs with a non-vanishing reward. 

For this class of agents, we proved that the probability $p(h_r)$ to observe a given rewarded history for a quantum-enhanced agent is equal to the one for a corresponding classical agent. As a consequence, a hybrid agent and its corresponding classical agent behave similarly. That is, quantum and classical agents with the same rewarded history $h_r$ follow the same policy and quantum and classical agents tend towards the same   average reward per epoch in the long run. 

Based on this result, we proved a quadratic speedup in learning for quantum-enhanced agents compared to their classical counterparts without sacrificing  the quality of the learned solution. Furthermore, we also analyzed  speedups which can be obtained with limited quantum hardware. We demonstrated that a quantum improvement can also be obtained if only a limited number of consecutive epochs can be performed coherently. 

A proof of principle experiment of a quantum-enhanced learning agent as discussed in this paper has been demonstrated experimentally with a nanophotonic quantum processor \cite{Saggio2021}. This concept of photonic quantum-enhanced learning agents can easily be expanded  to more advanced photonic architectures for reinforcement learning  \cite{Flamini_2020}.

In general, also classical learning agents, which do not obey the two criteria mentioned above, can be combined with quantum exploration. However, the behavior of such quantum-enhanced learning agents will differ from the behavior of the corresponding classical agents. For example, the probability for an epoch with vanishing reward is different for quantum and classical agents. As a consequence, policy updates based on epochs with vanishing reward might lead in the long run to different policies for quantum-enhanced agents and classical agents. Therefore, no general comparison between quantum-enhanced agents and classical agents  about the long term expected reward and learning time can be made in these cases. 

Many learning setups are luck-favoring in the sense that agents which received more rewards in the beginning  also receive more rewards on average in the future (compare \cite{Dunjko2016}). In general, quantum-enhanced agents will learn faster in such setups. However, the obtainable speedup will depend on the exact setup. 
In very rare occasions, an agent which observed fewer rewards in the beginning might perform better in the long run. We do not expect any quantum speedups in learning in such setups.

In the future, it is necessary to further study realistic learning setups based on general classical agents and faulty quantum hardware for example with the help of simulations and further numerical analyzes. In this way, more specific predictions about possible quantum improvements might be gained.   

\section*{Acknowledgments}
AH and SW thanks Hans Briegel, Vedran Dunjko, Davide Orsucci and Oliver Sefrin for fruitful discussions. AH acknowledges support from the Austrian Science Fund (FWF) through the project P 30937-N27.
SW acknowledges support from the Austrian Science Fund (FWF) through SFB BeyondC F7102.

\bibliographystyle{unsrtnat}
\bibliography{ref.bib}

\begin{thebibliography}{32}
\providecommand{\natexlab}[1]{#1}
\providecommand{\url}[1]{\texttt{#1}}
\expandafter\ifx\csname urlstyle\endcsname\relax
  \providecommand{\doi}[1]{doi: #1}\else
  \providecommand{\doi}{doi: \begingroup \urlstyle{rm}\Url}\fi

\bibitem[Dunjko and Briegel(2018)]{Dunjko2018}
Vedran Dunjko and Hans Briegel.
\newblock Machine learning \& artificial intelligence in the quantum domain: A
  review of recent progress.
\newblock \emph{Reports on Progress in Physics}, 81, 03 2018.
\newblock \doi{10.1088/1361-6633/aab406}.
\newblock URL \url{https://doi.org/10.1088/1361-6633/aab406}.

\bibitem[Biamonte et~al.(2017)Biamonte, Wittek, Pancotti, Rebentrost, Wiebe,
  and Lloyd]{biamonte2017}
Jacob Biamonte, Peter Wittek, Nicola Pancotti, Patrick Rebentrost, Nathan
  Wiebe, and Seth Lloyd.
\newblock Quantum machine learning.
\newblock \emph{Nature}, 549\penalty0 (7671):\penalty0 195--202, Sep 2017.
\newblock ISSN 1476-4687.
\newblock \doi{10.1038/nature23474}.
\newblock URL \url{https://doi.org/10.1038/nature23474}.

\bibitem[Sutton and Barto(1998)]{Sutton1998}
R.S. Sutton and A.G. Barto.
\newblock \emph{Reinforcement learning}.
\newblock The MIT Press, 1998.
\newblock ISBN 9780262193986.

\bibitem[Ciliberto et~al.(2018)Ciliberto, Herbster, Ialongo, Pontil, Rocchetto,
  Severini, and Wossnig]{Ciliberto2017}
Carlo Ciliberto, Mark Herbster, Alessandro~Davide Ialongo, Massimiliano Pontil,
  Andrea Rocchetto, Simone Severini, and Leonard Wossnig.
\newblock Quantum machine learning: a classical perspective.
\newblock \emph{Proceedings of the Royal Society A: Mathematical, Physical and
  Engineering Sciences}, 474\penalty0 (2209):\penalty0 20170551, 2018.
\newblock \doi{10.1098/rspa.2017.0551}.
\newblock URL
  \url{https://royalsocietypublishing.org/doi/abs/10.1098/rspa.2017.0551}.

\bibitem[Havl{\'i}cek et~al.(2019)Havl{\'i}cek, C{\'o}rcoles, Temme, Harrow,
  Kandala, Chow, and Gambetta]{Havlicek2019}
Vojtech Havl{\'i}cek, Antonio~D. C{\'o}rcoles, Kristan Temme, Aram~W. Harrow,
  Abhinav Kandala, Jerry~M. Chow, and Jay~M. Gambetta.
\newblock Supervised learning with quantum-enhanced feature spaces.
\newblock \emph{Nature}, 567\penalty0 (7747):\penalty0 209--212, 2019.
\newblock ISSN 1476-4687.
\newblock \doi{10.1038/s41586-019-0980-2}.
\newblock URL \url{https://doi.org/10.1038/s41586-019-0980-2}.

\bibitem[Schuld et~al.(2015)Schuld, Sinayskiy, and Petruccione]{Schuld2015}
Maria Schuld, Ilya Sinayskiy, and Francesco Petruccione.
\newblock An introduction to quantum machine learning.
\newblock \emph{Contemporary Physics}, 56\penalty0 (2):\penalty0 172--185,
  2015.
\newblock \doi{10.1080/00107514.2014.964942}.
\newblock URL \url{https://doi.org/10.1080/00107514.2014.964942}.

\bibitem[Adcock et~al.(2015)Adcock, Allen, Day, Frick, Hinchliff, Johnson,
  Morley-Short, Pallister, Price, and Stanisic]{adcock2015advances}
Jeremy Adcock, Euan Allen, Matthew Day, Stefan Frick, Janna Hinchliff, Mack
  Johnson, Sam Morley-Short, Sam Pallister, Alasdair Price, and Stasja
  Stanisic.
\newblock Advances in quantum machine learning, 2015.

\bibitem[Johannink et~al.(2019)Johannink, Bahl, Nair, Luo, Kumar, Loskyll,
  Ojea, Solowjow, and Levine]{johannink2019residual}
Tobias Johannink, Shikhar Bahl, Ashvin Nair, Jianlan Luo, Avinash Kumar,
  Matthias Loskyll, Juan~Aparicio Ojea, Eugen Solowjow, and Sergey Levine.
\newblock Residual reinforcement learning for robot control.
\newblock In \emph{2019 International Conference on Robotics and Automation
  (ICRA), Montreal, QC, Canada}, pages 6023--6029. IEEE, 2019.
\newblock \doi{10.1109/ICRA.2019.8794127}.
\newblock URL \url{https://doi.org/10.1109/ICRA.2019.8794127}.

\bibitem[Tjandra et~al.(2018)Tjandra, Sakti, and Nakamura]{tjandra2018sequence}
Andros Tjandra, Sakriani Sakti, and Satoshi Nakamura.
\newblock Sequence-to-sequence asr optimization via reinforcement learning.
\newblock In \emph{2018 IEEE International Conference on Acoustics, Speech and
  Signal Processing (ICASSP), Calgary, AB, Canada}, pages 5829--5833. IEEE,
  2018.
\newblock \doi{10.1109/ICASSP.2018.8461705}.
\newblock URL \url{https://doi.org/10.1109/ICASSP.2018.8461705}.

\bibitem[Komorowski et~al.(2018)Komorowski, Celi, Badawi, Gordon, and
  Faisal]{komorowski2018artificial}
Matthieu Komorowski, Leo~A. Celi, Omar Badawi, Anthony~C. Gordon, and A.~Aldo
  Faisal.
\newblock The artificial intelligence clinician learns optimal treatment
  strategies for sepsis in intensive care.
\newblock \emph{Nat. Med.}, 24\penalty0 (11):\penalty0 1716--1720, 2018.
\newblock \doi{10.1038/s41591-018-0213-5}.
\newblock URL \url{https://doi.org/10.1038/s41591-018-0213-5}.

\bibitem[Silver et~al.(2016)Silver, Huang, Maddison, Guez, Sifre, van~den
  Driessche, Schrittwieser, Antonoglou, Panneershelvam, Lanctot, Dieleman,
  Grewe, Nham, Kalchbrenner, Sutskever, Lillicrap, Leach, Kavukcuoglu, Graepel,
  and Hassabis]{Silver2016}
David Silver, Aja Huang, Chris~J. Maddison, Arthur Guez, Laurent Sifre, George
  van~den Driessche, Julian Schrittwieser, Ioannis Antonoglou, Veda
  Panneershelvam, Marc Lanctot, Sander Dieleman, Dominik Grewe, John Nham, Nal
  Kalchbrenner, Ilya Sutskever, Timothy Lillicrap, Madeleine Leach, Koray
  Kavukcuoglu, Thore Graepel, and Demis Hassabis.
\newblock Mastering the game of go with deep neural networks and tree search.
\newblock \emph{Nature}, 529\penalty0 (7587):\penalty0 484--489, 2016.
\newblock ISSN 1476-4687.
\newblock \doi{10.1038/nature16961}.
\newblock URL \url{https://doi.org/10.1038/nature16961}.

\bibitem[Jerbi et~al.(2021{\natexlab{a}})Jerbi, Trenkwalder, Poulsen~Nautrup,
  Briegel, and Dunjko]{Jerbi2021}
Sofiene Jerbi, Lea~M. Trenkwalder, Hendrik Poulsen~Nautrup, Hans~J. Briegel,
  and Vedran Dunjko.
\newblock Quantum enhancements for deep reinforcement learning in large spaces.
\newblock \emph{PRX Quantum}, 2:\penalty0 010328, Feb 2021{\natexlab{a}}.
\newblock \doi{10.1103/PRXQuantum.2.010328}.
\newblock URL \url{https://link.aps.org/doi/10.1103/PRXQuantum.2.010328}.

\bibitem[Paparo et~al.(2014)Paparo, Dunjko, Makmal, Martin-Delgado, and
  Briegel]{PhysRevX.4.031002}
Giuseppe~Davide Paparo, Vedran Dunjko, Adi Makmal, Miguel~Angel Martin-Delgado,
  and Hans~J. Briegel.
\newblock Quantum speedup for active learning agents.
\newblock \emph{Phys. Rev. X}, 4:\penalty0 031002, Jul 2014.
\newblock \doi{10.1103/PhysRevX.4.031002}.
\newblock URL \url{https://link.aps.org/doi/10.1103/PhysRevX.4.031002}.

\bibitem[Sriarunothai et~al.(2018)Sriarunothai, W\"{o}lk, Giri, Friis, Dunjko,
  Briegel, and Wunderlich]{Sriarunothai2018}
Th~Sriarunothai, S~W\"{o}lk, G~S Giri, N~Friis, V~Dunjko, H~J Briegel, and
  Ch~Wunderlich.
\newblock Speeding-up the decision making of a learning agent using an ion trap
  quantum processor.
\newblock \emph{Quantum Science and Technology}, 4\penalty0 (1):\penalty0
  015014, December 2018.
\newblock \doi{10.1088/2058-9565/aaef5e}.
\newblock URL \url{https://doi.org/10.1088/2058-9565/aaef5e}.

\bibitem[Jerbi et~al.(2021{\natexlab{b}})Jerbi, Gyurik, Marshall, Briegel, and
  Dunjko]{jerbi2021variational}
Sofiene Jerbi, Casper Gyurik, Simon Marshall, Hans~J. Briegel, and Vedran
  Dunjko.
\newblock Variational quantum policies for reinforcement learning,
  2021{\natexlab{b}}.

\bibitem[Ronagh(2019)]{Ronagh2019}
Pooya Ronagh.
\newblock Quantum algorithms for solving dynamic programming problems, 2019.

\bibitem[Crawford et~al.(2019)Crawford, Levit, Ghadermarzy, Oberoi, and
  Ronagh]{Crawford2018}
Daniel Crawford, Anna Levit, Navid Ghadermarzy, Jaspreet~S. Oberoi, and Pooya
  Ronagh.
\newblock Reinforcement learning using quantum boltzmann machines.
\newblock 2019.

\bibitem[Cornelissen(2018)]{Cornelissen2018}
Arjan Cornelissen.
\newblock Quantum gradient estimation and its application to quantum
  reinforcement learning, 2018.
\newblock URL
  \url{http://resolver.tudelft.nl/uuid:26fe945f-f02e-4ef7-bdcb-0a2369eb867e}.

\bibitem[Neukart et~al.(2018)Neukart, Von~Dollen, Seidel, and
  Compostella]{Neukart2018}
Florian Neukart, David Von~Dollen, Christian Seidel, and Gabriele Compostella.
\newblock Quantum-enhanced reinforcement learning for finite-episode games with
  discrete state spaces.
\newblock \emph{Frontiers in Physics}, 5:\penalty0 71, 2018.
\newblock ISSN 2296-424X.
\newblock \doi{10.3389/fphy.2017.00071}.
\newblock URL
  \url{https://www.frontiersin.org/article/10.3389/fphy.2017.00071}.

\bibitem[Casal{\'e} et~al.(2020)Casal{\'e}, Di~Molfetta, Kadri, and
  Ralaivola]{Casale2020}
Balthazar Casal{\'e}, Giuseppe Di~Molfetta, Hachem Kadri, and Liva Ralaivola.
\newblock Quantum bandits.
\newblock \emph{Quantum Machine Intelligence}, 2\penalty0 (1):\penalty0 11, Aug
  2020.
\newblock ISSN 2524-4914.
\newblock \doi{10.1007/s42484-020-00024-8}.
\newblock URL \url{https://doi.org/10.1007/s42484-020-00024-8}.

\bibitem[Saggio et~al.(2021)Saggio, Asenbeck, Hamann, Str{\"o}mberg, Schiansky,
  Dunjko, Friis, Harris, Hochberg, Englund, W{\"o}lk, Briegel, and
  Walther]{Saggio2021}
V.~Saggio, B.~E. Asenbeck, A.~Hamann, T.~Str{\"o}mberg, P.~Schiansky,
  V.~Dunjko, N.~Friis, N.~C. Harris, M.~Hochberg, D.~Englund, S.~W{\"o}lk,
  H.~J. Briegel, and P.~Walther.
\newblock Experimental quantum speed-up in reinforcement learning agents.
\newblock \emph{Nature}, 591\penalty0 (7849):\penalty0 229--233, Mar 2021.
\newblock ISSN 1476-4687.
\newblock \doi{10.1038/s41586-021-03242-7}.
\newblock URL \url{https://doi.org/10.1038/s41586-021-03242-7}.

\bibitem[Dunjko et~al.(2016)Dunjko, Taylor, and Briegel]{Dunjko2016}
Vedran Dunjko, Jacob~M. Taylor, and Hans~J. Briegel.
\newblock Quantum-enhanced machine learning.
\newblock \emph{Phys. Rev. Lett.}, 117:\penalty0 130501, Sep 2016.
\newblock \doi{10.1103/PhysRevLett.117.130501}.
\newblock URL \url{https://link.aps.org/doi/10.1103/PhysRevLett.117.130501}.

\bibitem[Hamann et~al.(2020)Hamann, Dunjko, and
  Wölk]{hamann2020quantumaccessible}
A.~Hamann, V.~Dunjko, and S.~Wölk.
\newblock Quantum-accessible reinforcement learning beyond strictly epochal
  environments, 2020.

\bibitem[Dunjko et~al.(2017)Dunjko, Liu, Wu, and Taylor]{dunjko2017exponential}
Vedran Dunjko, Yi-Kai Liu, Xingyao Wu, and Jacob~M. Taylor.
\newblock Exponential improvements for quantum-accessible reinforcement
  learning, 2017.

\bibitem[Mnih et~al.(2015)Mnih, Kavukcuoglu, Silver, Rusu, Veness, Bellemare,
  Graves, Riedmiller, Fidjeland, Ostrovski, Petersen, Beattie, Sadik,
  Antonoglou, King, Kumaran, Wierstra, Legg, and Hassabis]{Mnih2015}
Volodymyr Mnih, Koray Kavukcuoglu, David Silver, Andrei~A. Rusu, Joel Veness,
  Marc~G. Bellemare, Alex Graves, Martin Riedmiller, Andreas~K. Fidjeland,
  Georg Ostrovski, Stig Petersen, Charles Beattie, Amir Sadik, Ioannis
  Antonoglou, Helen King, Dharshan Kumaran, Daan Wierstra, Shane Legg, and
  Demis Hassabis.
\newblock Human-level control through deep reinforcement learning.
\newblock \emph{Nature}, 518\penalty0 (7540):\penalty0 529--533, Feb 2015.
\newblock ISSN 1476-4687.
\newblock \doi{10.1038/nature14236}.
\newblock URL \url{https://doi.org/10.1038/nature14236}.

\bibitem[Briegel and De~las Cuevas(2012)]{Briegel2012}
Hans~J. Briegel and Gemma De~las Cuevas.
\newblock Projective simulation for artificial intelligence.
\newblock \emph{Sci. Rep.}, 2:\penalty0 400, May 2012.
\newblock ISSN 2045-2322.
\newblock URL \url{http://dx.doi.org/10.1038/srep00400}.

\bibitem[Maschler et~al.(2020)Maschler, Solan, and Zamir]{maschler2020game}
M.~Maschler, E.~Solan, and S.~Zamir.
\newblock \emph{Game Theory}.
\newblock Cambridge University Press, 2020.
\newblock ISBN 9781108659956.
\newblock URL \url{https://books.google.at/books?id=DtjrDwAAQBAJ}.

\bibitem[Grover(1998)]{Grover1998}
Lov~K. Grover.
\newblock Quantum computers can search rapidly by using almost any
  transformation.
\newblock \emph{Phys. Rev. Lett.}, 80:\penalty0 4329, 1998.
\newblock \doi{10.1103/PhysRevLett.80.4329}.
\newblock URL \url{https://link.aps.org/doi/10.1103/PhysRevLett.80.4329}.

\bibitem[Boyer et~al.(1998)Boyer, Brassard, Høyer, and Tapp]{Boyer1998}
Michel Boyer, Gilles Brassard, Peter Høyer, and Alain Tapp.
\newblock Tight bounds on quantum searching.
\newblock \emph{Fortschritte der Physik}, 46\penalty0 (4-5):\penalty0 493--505,
  1998.
\newblock \doi{10.1002/3527603093.ch10}.

\bibitem[Yoder et~al.(2014)Yoder, Low, and Chuang]{PhysRevLett.113.210501}
Theodore~J. Yoder, Guang~Hao Low, and Isaac~L. Chuang.
\newblock Fixed-point quantum search with an optimal number of queries.
\newblock \emph{Phys. Rev. Lett.}, 113:\penalty0 210501, Nov 2014.
\newblock \doi{10.1103/PhysRevLett.113.210501}.
\newblock URL \url{https://link.aps.org/doi/10.1103/PhysRevLett.113.210501}.

\bibitem[Flamini et~al.(2020)Flamini, Hamann, Jerbi, Trenkwalder, Nautrup, and
  Briegel]{Flamini_2020}
Fulvio Flamini, Arne Hamann, Sofi{\`{e}}ne Jerbi, Lea~M Trenkwalder,
  Hendrik~Poulsen Nautrup, and Hans~J Briegel.
\newblock Photonic architecture for reinforcement learning.
\newblock \emph{New Journal of Physics}, 22\penalty0 (4):\penalty0 045002, apr
  2020.
\newblock \doi{10.1088/1367-2630/ab783c}.
\newblock URL \url{https://doi.org/10.1088%2F1367-2630%2Fab783c}.

\bibitem[Melnikov et~al.(2018)Melnikov, Makmal, and
  Briegel]{MelnikovMakmalBriegel2018}
Alexey~A. Melnikov, Adi Makmal, and Hans~J. Briegel.
\newblock Benchmarking projective simulation in navigation problems.
\newblock \emph{IEEE Access}, 6:\penalty0 64639--64648, 2018.
\newblock \doi{10.1109/ACCESS.2018.2876494}.
\newblock URL \url{https://doi.org/10.1109/ACCESS.2018.2876494}.

\end{thebibliography}

\appendix
\section*{Appendix}

\section{Turnover from quantum to classical search}\label{appendix:Q_max}
A classical agent can observe a reward with probability $Q$, \eq{eq:win_prob} in every epoch. As a result, the average reward  obtained by classical agents following the policy $\hat{\Pi}$ is given by 
\be
\langle r \rangle_c = \sum\limits_{\lbrace \vec{a} \rbrace} r(\vec{a})
\hat{\Pi}(\vec{a})=\Bar{r}_c \; Q.
\ee
Here, we introduced the average reward of a winning sequence
\be
\Bar{r}_c=\sum\limits_{\lbrace \vec{a}| r(\vec{a})>0 \rbrace} r(\vec{a})
\hat{\Pi}(\vec{a})/Q
\ee
for classical agents, which is equal to the average reward conditioned on observing a reward $r>0$.

A quantum-enhanced agent uses $\alpha_o k$ epochs to perform amplitude amplification to determine a possible rewarded sequence of actions. The agent will receive in an additional consecutive epoch a reward with probability \cite{Boyer1998}

    \begin{align}
        G(Q,k)&=\sin^2\left((2k+1)\Theta(Q) \right) \\
        \Theta(Q) &= \arcsin{\sqrt{Q}}.
    \end{align}
Thus, quantum-enhanced agents  receive on average a reward of     
 \be
 \langle r \rangle_q=\Bar{r}_q\frac{G(Q,k)}{\alpha_o k+1}
 \ee
 where $\Bar{r}_q$ is the  average reward of a winning sequence
of a quantum-enhanced agent. Due to \theo{theorem:same_rewarded_histories}, we find $\Bar{r}_q=\Bar{r}_c = \Bar{r}$. As a consequence, a learning agent with a winning probability $Q$ will receive on average more rewards when performing quantum exploration for $k$ epochs if
\be
Q < \frac{\sin^2\left[(2k+1)\arcsin(\sqrt{Q}) \right] }{\alpha_o k +1}.
\ee
The inequality can only be solved numerically. For $\alpha_o=1$ and $k=1$ we find that a quantum-enhanced agent finds on average more rewards if $Q<Q_\text{max}\approx 0.3964$.

\section{\theo{theorem:same_rewarded_histories}}\label{appendix:same_rewarded_histories}
In the following, we give a detailed proof of \theo{theorem:same_rewarded_histories}. Please keep in mind that we consider in this paper
 only agents, where all policy updates are completely determined by their rewarded history. Therefore, we introduce $\hat{\Pi}_j$ as the policy of an agent in the $j$-th interval starting after the $j$-th rewarded epoch and ending with the $j+1$-th rewarded epoch. As a result, the probability to play the sequence of actions $\vec{a}$ in an epoch in the interval $j$ is given by $\hat{\Pi}_j(\vec{a})$.

The probability to get no reward in an epoch in the $j$-th interval is given by
\be
\sum_{\lbrace \vec{a}'| r(\vec{a}')=0 \rbrace} \hat{\Pi}_j(\vec{a'}) = 1-Q_j
\ee
with $Q_j$,\eq{eq:win_prob}, being the reward probability in this interval.
The probability $p_j(t_j,\vec{a}_j)$ that the $j$-th interval contains $t_j$ epochs and ends with the action sequence $\vec{a}_j$ with $r(\vec{a}_j)>0$ is given by the probability to play $t_j-1$ epochs without getting any reward and play then the action sequence $\vec{a}_j$ leading to 
\be
p_j(t_j,\vec{a}_j)=(1-Q_j)^{t_j-1} \hat{\Pi}_j(\vec{a}_j).\label{eq:p_j_app}
\ee

As a consequence, the probability $p_j(\vec{a}_j)$ that playing the action sequence $\vec{a}_j$ will lead to the $j$-th observed reward is given by
\be
\begin{split}
p_j(\vec{a}_j)&=\sum\limits_{t_j=1}^\infty p_j(t_j,\vec{a}_j)\\
&=\frac{1}{Q_j} \hat{\Pi}_j(\vec{a}_j), \label{eq:prop_a_j}
\end{split}
\ee
where we used the geometric series in the second step.
The probability to observe a given rewarded history $h_r$ 
can be expressed  with the help of the conditional probability $p(h_r|\vec{a}_0)$, denoting the probability to  observe $h_r$ if $\vec{a}_0$ was observed as the first rewarded action sequence via
\begin{align}
    p(h_r)&= p(h_r|\vec{a}_0)p_0(\vec{a}_0).
\end{align}
The same considerations hold for all other time intervals, leading together with \eq{eq:prop_a_j} to
\begin{align}
    p(h_r)&=\prod_{j=0}^{J-1} \frac{\hat{\Pi}_j(\vec{a}_j)}{Q_j}.\label{eq:prob_history},
\end{align}
where we assumed that the rewarded history $h_r$ contains $J$ rewarded epochs. 

Amplitude amplification \cite{Boyer1998, Grover1998} enhances $Q_j$ but preserves the ratio $\hat{\Pi}_j(\vec{a}_j)/Q_j$. The classical and the hybrid agent start with the same policy $\hat{\Pi}_0$. As a consequence, their probability to observe $\vec{a}_0$ as the first rewarded action sequence is equal leading to identical probabilities $p(h_r)$ for rewarded histories of length $J=1$. If a quantum and a classical agent played the same $\vec{a}_0$ as first rewarded action sequences, they follow the same policy $\hat{\Pi}_1$ in the next interval leading together with \eq{eq:prob_history} to identical distributions for rewarded histories.

\section{Average learning times}\label{appendix:proof_learning_time}
In the following, we summarize more detailed discussions concerning the proof of \theo{theorem:learning_time}.

The expected learning time $\langle T(h_r) \rangle$ for an agent with a given rewarded history can be expressed by 
\be
\langle T(h_r) \rangle = \sum_{j=0}^{J(h_r)-1} \langle t_j(h_r) \rangle,
\ee
because the behavior of the here considered agents depends solely on the rewards an agent has found so far.

The probability $p_j(t_j,\vec{a}_j)$ that the duration of interval $j$ is given by $t_j$ epochs and that interval $j$ ends with the rewarded action sequence $\vec{a}_j$ is given by \eq{eq:p_j_app}.
The probability that the duration of interval $j$ is given by $t_j$ conditioned on that the agent's rewarded history is given by $h_r$ is therefore determined by

\be
p_j(t_j|h_r)=\frac{p_j(t_j,\vec{a}_j)}{p_j(\vec{a}_j)}=Q_j (1-Q_j)^{t_j-1}.
\ee
 As a consequence, the average interval time $\langle t_j(h_r) \rangle$ of an agent with rewarded history $h_r$ is given by
\be
\langle t_j(h_r) \rangle = \sum_{\tau=1}^\infty \tau p_j(\tau|h_r)=\frac{1}{Q_j}
\ee
as used in \eq{eq:t_j(h)}.

In the main text, we then express the average learning time $\langle T\rangle$ as an average of the learning times $\langle T(h_r) \rangle$ for different given rewarded histories in \eq{eq:T_average}. This equation follows from the following considerations.

Let $p(T,h_r)$ be the probability that an agent has learned after $T$ epochs and has observed the rewarded history $h_r$ and let $H_r$ be the set of all possible rewarded histories. Then, the average learning time is determined by
\be
\langle T \rangle = \sum_{T=1}^\infty \sum_{h_r\in H_r} T p(T,h_r).
\ee
The average learning time $\langle T(h_r)\rangle$ of an agent conditioned on observing $h_r$ is given by
\be
\langle T(h_r) \rangle = \sum_{T=1}^\infty  T p(T|h_r)
\ee
with $p(T|h_r)=p(T,h_r)/p(h_r)$. As a consequence, we find
\be
\langle T \rangle = \sum_{T=1}^\infty \sum_{h_r\in H_r} T p(T|h_r)p(h_r)= \sum_{h_r\in H_r}\langle T(h_r)\rangle p(h_r).
\ee

\section{Binary tree environment}\label{appendix:binary_tree_env}
In this section, we describe the binary tree environment  used as an example environment for \fig{fig:learning-curve}. Additionally, we introduce a learning agent based on t-PS \cite{Flamini_2020} simplified and reduced to the essentials for this binary tree environment. For a more detailed introduction of projective simulation or  t-PS we recommend \cite{Briegel2012} and \cite{Flamini_2020}.

\begin{figure*}
    \centering
    \includegraphics[width=0.75\linewidth]{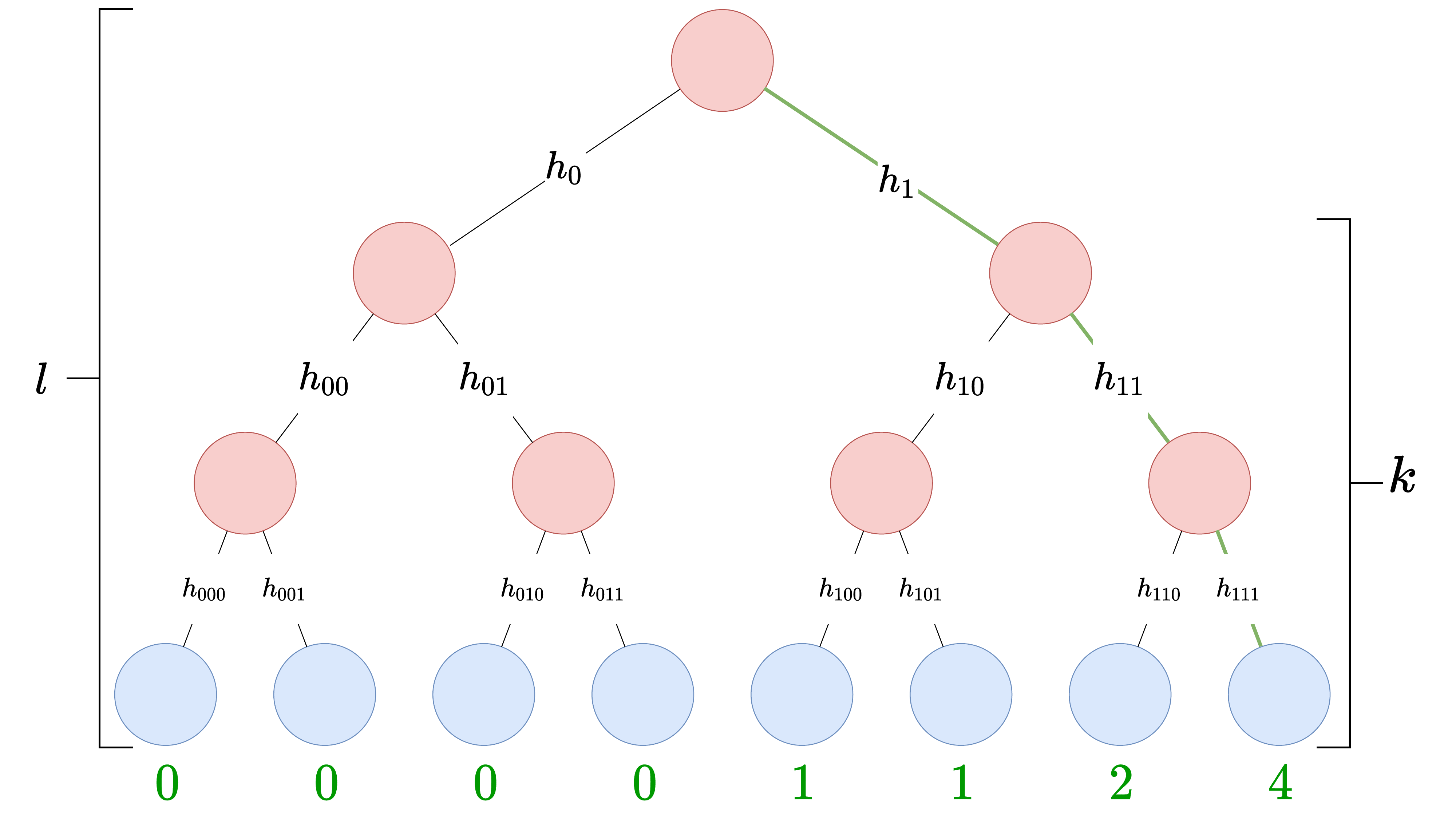}
    \caption{Binary tree environment with $l=3$ layers. The correct sequence (green) is rewarded with a reward $r=2^k$, leaving the correct sequence after $x$ correct decisions leads to an exponentially decaying reward with $r=\lfloor2^{(k+x-l)}\rfloor$ (green numbers). 
    Each binary choice is connected to a so-called $h$-value determining the probability to choose this option. 
    At the end of each epoch (after $l$ decisions) all $h$-values along the chosen sequence are increased by the obtained reward.}
    \label{fig:binary_tree_env}
\end{figure*}

In the binary tree environment, see \fig{fig:binary_tree_env}, an agent has to perform $l$ sequential binary decisions in an epoch. Then, a reward is issued and a new epoch starts. 
One sequence of action is marked as the correct sequence and rewarded with a reward $r=2^k$. Leaving the correct sequence after $x$ correct decisions leads to an exponentially decaying reward of $r=\lfloor2^{(k+x-l)}\rfloor$. The number of rewarded action sequences ($r>0$) relative to the complete number of possible action sequences reduces exponentially with $l-k$. On the other hand, the reward difference within the rewarded subspace grows  exponentially with $k$. 

This environment inhabits two important features making it an ideal example for comparing the here described hybrid agent with its classical counterpart:(i) It is hard to find a rewarded action sequence leading to a big possible speedup for the hybrid agent. (ii) Classical optimization within the rewarded subspace leads to higher rewards, allowing a comparison of the quality of the found solution between the hybrid agent and its classical counterpart.

To complete the example, we will need an agent interacting with this environment. As this paper does not focus on how to construct a good classical agent, we will use just a simple agent suited for this environment. We would like to emphasize that more advanced agents could be used, too. Our simple agent assigns an $h$ value to each possible choice.  Each of the $l$ binary decisions is decided by  a random decision governed by a softmax policy.  the $y$-th decision of an epoch is thus governed by the probabilities
\be
p_{i_y} = 0.5 + 0.5\tanh{\left[\beta (h_{i_1,\cdots,i_y} - h_{i_1,\cdots\lnot i_y})\right]},
\ee
with $i_z\in\lbrace 0,1\rbrace\,\forall\, 1\leq z\leq l$ and
 $\lnot i$ denoting the alternative  choice to $i$. The $h$ values are initialized by the same value, which is irrelevant because the policy only depends on the difference between the $h$-values. If a rewarded sequence $\vec{i}=(i_1,\cdots,i_l)$ was chosen, all $h$ values of the sequence are increased by the obtained reward $r$ via
 \be
 h_{i_1,\cdots,i_y}\rightarrow  h_{i_1,\cdots,i_y}+r\quad 1\leq y\leq l.
 \ee
\end{document}